\documentclass[11pt]{article}
\usepackage{amsmath}
\makeatletter
\begin{document}
\begin{center}
\huge{\textbf{The Extended Wronskian Determinant Approach and the
Iterative Solutions of One-Dimensional Dirac Equation}}
\end{center}
\begin{center}
Ying Xu, Meng Lu\medskip

\textit{Department of Physics, Fudan University, Shanghai, 200433,
P.~R.~China}\medskip

Ru-Keng Su\medskip

\textit{China Center of Advanced Science and Technology (World
Laboratory),} \\\textit{P.~O.~Box 8730, Beijing 100080,
P.~R.~China}; \\\textit{Department of Physics, Fudan University,
Shanghai, 200433, P.~R.~China}\medskip
\end{center}

\begin{center}
\textbf{\large Abstract}
\end{center}

An approximation method, namely, the Extended Wronskian
Determinant Approach, is suggested to study the one-dimensional
Dirac equation. An integral equation which can be solved by
iterative procedure to find the wave functions is established. We
employ this approach to study the one-dimensional Dirac equation
with one-well potential, and give the energy levels and wave
functions up to the first order iterative approximation. For
double-well potential, the energy levels up to the first order
approximation are given.


\newline
\newline
\newline
\newline
\newline
\newline
\newline
\newline
\newline
\newline
\newline
\newline
\begin{flushright}
Typeset using \LaTeX
\end{flushright}
\pagebreak
\section{Introduction}
Since the majority of problems in quantum mechanics lead to very
complex equations and cannot be solved exactly, the approximation
methods for solving Schr\"odinger equation are very important for
practical use. Recently, Friedberg, Lee and Zhao(FLZ) suggested a
new method to decide the wave functions and energy levels of
Schr\"odinger equation by quadrature along a single
trajectory~[1-3]. Giving a good trial wave function $\Psi$ for
one-dimensional(1D) Schr\"odinger equation with potential $V$,
\begin{equation}
  -\frac{1}{2}\frac{\mathrm{d}^2\Phi}{\mathrm{d}x^2}+V\Phi=E\Phi
\end{equation}
they considered a perturbation equation
\begin{equation}
  -\frac{1}{2}\frac{\mathrm{d}^2\Psi}{\mathrm{d}x^2}+(V+w)\Psi=(E+\varepsilon)\Psi
\end{equation}
where $w$ is the perturbation potential and $\varepsilon$ the
energy shift. By means of the Green's function, they constructed
an integral equation and solved it by iterative procedure to find
the true ground wave function and the corresponding energy level.
The integral equation for wave function $\Psi$ reads
\begin{equation}
  \Psi=\Phi-2\Phi\int_x^{+\infty}\Phi^{-2}(y)dy~\int_y^{+\infty}\Phi(z)(w-z)\Psi(z)dz
\end{equation}
and the corresponding energy is
\begin{equation}
  \varepsilon=\frac{\displaystyle\int_0^{+\infty}\Phi w\Psi \mathrm{d}x}{\displaystyle\int_0^{+\infty}\Phi\Psi \mathrm{d}x}
\end{equation}
Employing a double-well potential\cite{3} and other potentials[4,
5] they gave the ground state energy up to the first order
approximation of iterative procedure.

On the other hand, as is well known, the behavior of 1D
Schr\"odinger equation is determined  completely by its
corresponding Wronskian determinant~[6]. Therefore, a Wronskian
Determinant Approach(WDA) is suggested by us~[7] to study the
energy levels and the wave functions of 1D Schr\"odinger equation.
We established an integral equation and employed this equation to
study the same problem of double-well potential. A series of
expansion of ground state energy up to the second order
approximation of iterative procedure is given. We found that the
series of energy given by FLZ method and the WDA up to the order
of $O(g^{-1})$ where $g$ is the strong coupling constant are the
same. This result confirms that the WDA is a successful method to
investigate the 1D Schr\"odinger equation.

The objective of this paper is to extend the WDA to study the 1D
Dirac equation. Based on general considerations and an extended
Wronskian determinant, we establish an integral equation for Dirac
wave function and give the ``resonant condition" to determine the
energy level. Our approach may be considered as a relativistic
generalization of the WDA. We will show the formulae of our
Extended Wronskian Determinant Approach(EWDA) in next section. In
section III, as an example, we will use EWDA to study the 1D Dirac
equation with perturbative one-well potential. We will solve the
integral equation by iterative procedure up to the first order to
find the energy shifts and the wave functions. Since the
double-well potential for Schr\"odinger equation and, in
particular, for Dirac equation has wide applications in physics,
for example, soliton bag model, instantons and anti-instantons and
etc[8-12], we will address the energy levels and the wave
functions for double-well potential in section IV. The last
section is a summary.
\pagebreak
\section{The Extended Wronskian Determinant Approach}
The 1D Dirac equation reads
\begin{equation}\label{21}
\left( \hat{\alpha}\hat{p}+\hat{\beta}m\right) \psi \left(
y\right) =\left( E-V\right) \psi \left( y\right)
\end{equation}
where $\psi =\binom{u}{v}$ is a 2-component spinor. Choose $\hat{\alpha}=%
\hat{\sigma}_{2}$, $\hat{\beta}=\hat{\sigma}_{3}$ and define
\begin{equation}\label{22}
\mathbf{\hat{A}}\equiv\left(
\begin{matrix}
0 & m+E-V \\
m-E+V & 0
\end{matrix}
\right)
\end{equation}
then Eq.~(\ref{21}) could be rewritten as
\begin{equation}\label{23}
\psi ^{\prime }\left( y\right) -\mathbf{\hat{A}}\psi \left(
y\right)=0
\end{equation}
Eq.~(\ref{23}) is called the homogeneous equation in our approach,
We need its specific solution as a zeroth order approximation to
construct the solution of the perturbation equation. Assume a
perturbative potential $w$ is added to Eq.~(\ref{21})
\begin{equation}\label{24}
\mathcal{V}=V+w
\end{equation}
and the eigenenergy is shifted to
\begin{equation}\label{25}
\mathcal{E}=E+\varepsilon
\end{equation}
We get the perturbation equation
\begin{equation}\label{26}
\psi ^{\prime }\left( y\right)-\mathbf{\hat{A}}\psi =\left(
\varepsilon -w\right) \mathbf{\hat{f}}\psi
\end{equation}
where
\begin{equation}\label{27}
\mathbf{\hat{f}}\equiv\left(
\begin{matrix}
0 & 1 \\
-1 & 0
\end{matrix}
\right)
\end{equation}
Eq.~(\ref{26}) is called an inhomogeneous equation and the
inhomogeneous term is $\left( \varepsilon -w\right)
\mathbf{\hat{f}}\psi$ in our approach.

As is well known, in ordinary differential equation theory, with a
specific solution of a homogeneous differential equation given, we
can try to find another linearly independent solution of this
equation and the general solutions of the corresponding
inhomogeneous differential equation through Wronskian determinant
calculation. This procedure can also be employed to discuss the
solution of 1D Dirac equation.

Suppose $\psi _{1}={\binom{u_{1}}{v_{1}}}$ is a specific solution
of Eq.~(\ref{23}). It is a normalized bound state and satisfies
the boundary condition \[ \lim_{y \rightarrow \pm \infty}\psi_1=
0.\]Now we construct another solution $\psi
_{2}={\binom{u_{2}}{v_{2}}}$ which is a linearly independent
solution with $\psi_1$ by using the extended Wronskian determinant
approach. Obviously we have
\begin{equation}\label{28}
\left(
\begin{matrix}
v_{2} & -u_{2}
\end{matrix}
\right) \psi _{1}^{\prime }-\left(
\begin{matrix}
v_{1} & -u_{1}
\end{matrix}
\right) \psi _{2}^{\prime }=\left(
\begin{matrix}
v_{2} & -u_{2}
\end{matrix}
\right)\mathbf{\hat{A}}\psi _{1}-\left(
\begin{matrix}
v_{1} & -u_{1}
\end{matrix}
\right) \mathbf{\hat{A}}\psi _{2}
\end{equation}
A straightforward calculation can show the right hand side of Eq.~
(\ref{28}) be zero, we get
\begin{equation}\label{29}
\left( u_{1}v_{2}-u_{2}v_{1}\right) ^{\prime }=0
\end{equation}
Define a matrix $\mathbf{\hat{U}}$ as
\begin{equation}\label{999}
\mathbf{\hat{U}}\equiv\left(
\begin{matrix}
u_{1} & u_{1} \\
v_{1} & v_{2}
\end{matrix}
\right)
\end{equation}
We find
\begin{equation}\label{30}
\mathrm{det}\mathbf{\hat{U}}=C\left( \mathit{Const.}\right)
\end{equation}
Here $\mathrm{det}\mathbf{\hat{U}}$ is called an extended
Wronskian determinant. The constant $C$ is nonzero because
$\psi_1$ and $\psi_2$ are linearly independent. Since $\psi_1$
satisfies the boundary condition \[ \lim_{y \rightarrow \pm
\infty}\psi_1= 0,\] we get that $\psi_2$ must be divergent when
$y\rightarrow\pm\infty$ because of Eq.~(\ref{30}). Eqs.~(\ref{30})
and (\ref{23}) yield
\begin{equation}\label{31}
u_{1}v_{2}-\frac{1}{m-E+V}v_{2}^{\prime }v_{1}=C
\end{equation}
Assume $v_2(y)$ has the form $v_{2}\left( y\right)=h\left(
y\right) v_{1}\left( y\right)$ where $h(y)$ is an unknown function
which can be solved from Eqs.~(\ref{31}) and (\ref{23}), we find
\begin{equation}\label{32}
h(y)=-C\int \mathrm{d}y\frac{m-E+V}{v_{1}^{2}}
\end{equation}
and
\begin{equation}\label{33}
v_{2}=-C v_{1}\int \mathrm{d}y\frac{m-E+V}{v_{1}^{2}}
\end{equation}
Similar procedure can be used to construct $u_2$, we finally
obtain
\begin{equation}\label{96}
\psi_{2}=\left( \begin{matrix}
u_{2}\\
v_{2}
\end{matrix} \right) = - C \left(
\begin{matrix}
\displaystyle u_{1}\int \mathrm{d}y\frac{-m-E+V}{u_{1}^{2}} \\[12pt]
\displaystyle v_{1}\int \mathrm{d}y\frac{+m-E+V}{v_{1}^{2}}
\end{matrix}\right)
\end{equation}
$\psi_2$ dose not satisfy the boundary condition of $\psi_1$,
because of the uniqueness theorem.

Now we are in a position to find the solution
$\psi={\binom{u}{v}}$ of the corresponding inhomogeneous
perturbation equation (\ref{26}). Suppose $\psi$ has the form
\begin{equation}\label{34}
\psi=\left(
\begin{matrix}
u\\
v
\end{matrix}\right)=\mathbf{\hat{U}}
\left( y \right) g\left( y \right)
\end{equation}
where $g(y)$ is a column matrix. Substituting Eqs.~(\ref{34}) and
(\ref{999}) into Eq.~(\ref{26}) we obtain
\begin{equation}\label{35}
g^{\prime}=(\varepsilon-w)\mathbf{\hat{U}}^{-1}\mathbf{\hat{f}}\mathbf{\hat{U}}g
\equiv \mathbf{\hat{B}} g
\end{equation}
where $\mathbf{\hat{U}}^{-1}$ is the inverse matrix of
$\mathbf{\hat{U}}$. A simple calculation can show that
\begin{equation}\label{36}
\mathbf{\hat{B}}\left( \varepsilon-w, y
\right)\equiv(\varepsilon-w)\mathbf{\hat{U}}^{-1}\mathbf{\hat{f}}\mathbf{\hat{U}}=
\frac{1}{C}\left( \varepsilon-w \right) \left(
\begin{matrix}
u_{1}u_{2}+v_{1}v_{2} & u_{2}^{2}+v_{2}^{2}\\[9pt]
-(u_{1}^{2}+v_{1}^{2}) & -(u_{1}u_{2}+v_{1}v_{2})
\end{matrix}\right)
\end{equation}
and the wave function
\begin{equation}\label{98}
\psi=\mathbf{\hat{U}}\int \mathrm{d}y
\mathbf{\hat{B}}g=\mathbf{\hat{U}}\int
\mathrm{d}y(\varepsilon-w)\mathbf{\hat{U}}^{-1}\mathbf{\hat{f}}\psi
\end{equation}
or, explicitly,
\begin{equation}\label{0}
\left(
\begin{matrix}
u\\
v
\end{matrix}\right)=\left(
\begin{matrix}
u_{1} & u_{2}\\
v_{1} & v_{2}
\end{matrix}\right)
\int \mathrm{d}y \frac{1}{C}(\varepsilon-w)\left(
\begin{matrix}
v_{2} & -u_{2}\\
-v_{1} & u_{1}
\end{matrix}\right)\left(
\begin{matrix}
0 & 1\\
-1 & 0
\end{matrix}\right)\left(
\begin{matrix}
u\\
v
\end{matrix}\right)
\end{equation}
Eq.~(\ref{0}) is an integral equation and can be solved by
iterative procedure step by step. It can be proven that the choice
of the constant $C$ does not affect on our result remarkably. For
simplicity, we choose $C=1$ in the following calculations. The
integration bound of Eq.~(\ref{0}) is determined by the boundary
condition and the normalized condition. Based on Eq.~(\ref{35}),
we obtain an iterative sequence of $g$ as
\begin{equation}\label{37}\left\{\begin{aligned}
&g^{(0)}\left(\varepsilon-w,y\right)=g\left(\varepsilon-w,a\right)\\
&g^{(1)}\left(\varepsilon-w,y\right)=g\left(\varepsilon-w,a\right)+\int_a^y
\mathrm{d}y^{\prime}\mathbf{\hat{B}}\left(\varepsilon-w,y^{\prime}\right)g^{
(0)}\left(\varepsilon-w,y^{\prime}\right)\\
& \cdots \\
&g^{(n)}\left(\varepsilon-w,y\right)=g\left(\varepsilon-w,a\right)+\int_a^y
\mathrm{d}y^{\prime}\mathbf{\hat{B}}\left(\varepsilon-w,y^{\prime}\right)g^{
(n-1)}\left(\varepsilon-w,y^{\prime}\right)\  {}
\end{aligned}\right.
\end{equation}
The corresponding formal solution of wave function $\psi$ is given
by Eq.~(\ref{34}).

Now we turn to discuss the energy shift $\varepsilon$. We get
\begin{equation}\label{38}
  \varepsilon=\left(
\begin{matrix}
u\\
v
\end{matrix}\right)=
\left(
\begin{matrix}
\displaystyle u_1g_1+u_2g_2\\
\displaystyle v_1g_1+v_2g_2
\end{matrix}\right)
\end{equation}
from Eq.~(\ref{34}), where $g=\left(\begin{matrix}
g_1\\
g_2\end{matrix}\right)$. Noting that the boundary conditions for
$\psi_1$ and $\psi_2$ are
\begin{equation}\nonumber
\lim_{y\rightarrow\pm\infty}\psi_1=\lim_{y\rightarrow\pm\infty}
\left(\begin{matrix}
u_1\\
v_1\end{matrix}\right)=0
\end{equation}
and
\begin{equation}\nonumber
\lim_{y\rightarrow\pm\infty}\psi_2=\lim_{y\rightarrow\pm\infty}
\left(\begin{matrix}
u_2\\
v_2\end{matrix}\right)=\infty
\end{equation}
respectively, we have
\begin{equation}\label{40}
  \lim_{y\rightarrow\pm\infty}g_2=0
\end{equation}
According to Eqs. (\ref{35}) and (\ref{36}) we find
\begin{equation}\label{41}
  g_2(\varepsilon-w, y)=g_2(\varepsilon-w,
  a)+\frac{1}{C}\int_a^y\mathrm{d}y~(\varepsilon-w)[-(u_1^2+v_1^2)g_1-(u_1u_2+v_1v_2)g_2]
\end{equation}
By means of Eq.~(\ref{40}), we get
\begin{equation}\label{42}
\int_{-\infty}^{+\infty}\mathrm{d} y\left(\varepsilon-w \right)
\left[\left(u_{1}^{2}+v_{1}^{2}\right)g_{1}+\left(u_{1}u_{2}+v_{1}v_{2}\right)g_{2}\right]=0
\end{equation}
Eq.~(\ref{42}) is called the ``resonant condition". The energy
shift reads
\begin{equation}\label{1}
\varepsilon=\frac {\displaystyle\int_{-\infty}^{+\infty}\mathrm{d}
yw
\Big[\left(u_{1}^{2}+v_{1}^{2}\right)g_{1}+\left(u_{1}u_{2}+v_{1}v_{2}\right)g_{2}\Big]}
{\displaystyle\int_{-\infty}^{+\infty}\mathrm{d} y
\Big[\left(u_{1}^{2}+v_{1}^{2}\right)g_{1}+\left(u_{1}u_{2}+v_{1}v_{2}\right)g_{2}\Big]}
\end{equation}
Eqs.~(\ref{0}), (\ref{37}), (\ref{1}) as well as the normalization
condition of $\psi$ are the basic formulae in EWDA. The zeroth
order wave function can be taken as $\psi^{(0)}=\psi_1$ when $w$
is a perturbation potential. Starting from $\psi^{(0)}$, we can
find the wave function and the energy level up to any order which
we need in principle by iterative procedure.
\pagebreak
\section{One-Well Potential}
\subsection{Square-Well Potential}%
As an application and example, now we study how to find the bound
state wave functions and the energy levels of 1D Dirac equation
with one well-potential. As a homogeneous equation, we study the
square-well potential first.

The 1D Dirac equation with square-well potential
\begin{equation}\label{3}
V\left( y \right) =-V_{0}\theta \left( y+\frac {a}{2}
\right)\theta \left(-y+\frac {a}{2} \right)
\end{equation}
where $\theta \left( y\right)$ is the step function
\begin{equation}\theta\left(y\right)=\left\{\begin{aligned}
&1\qquad(y>0)\\
&0\qquad(y<0)\quad
\end{aligned}\right.
\end{equation}
(Fig.~1) can be solved exactly~[13]. The bound state solution
$\psi_1$ is
\begin{equation}\label{4}
\begin{aligned}\psi_{1}=
\left( \begin{matrix} u_{1}\\[5pt]
v_{1}
\end{matrix}\right)=&\left( \begin{matrix} 1\\[5pt]
\frac {m-E}{\kappa}
\end{matrix}\right)u_{A}e^{\kappa y}\theta \left(-y-\frac
{a}{2}\right)\\[5pt]
&+\Bigg [ \left( \begin{matrix} \cos py\\[5pt]
\frac {m-E-V_{0}}{p} \sin py
\end{matrix}\right)u_{B}+\left( \begin{matrix} \sin py\\[5pt]
-\frac {m-E-V_{0}}{p} \cos py
\end{matrix}\right)u_{C}\Bigg]\theta \left(y+\frac {a}{2}\right)\theta \left(-y+\frac
{a}{2}\right)\\[5pt]
&+\left( \begin{matrix} 1\\[5pt]
-\frac {m-E}{\kappa}
\end{matrix}\right)u_{D}e^{-\kappa y}\theta \left(y-\frac {a}{2}\right)
\end{aligned}\end{equation}
in which
\begin{equation}
  \kappa=\sqrt {m^{2}-E^{2}}\qquad\qquad\quad \left( -m<E<m \right)
\end{equation}
\begin{equation}
  p=\sqrt {\left(E+V_{0} \right)^{2}-m^{2}}\qquad \left( m-V_{0}<E \right)
\end{equation}
The coefficients $u_{A}, u_{B}, u_{C}$ and $u_{D}$ in one case are
given by
\begin{equation}
  u_{A}=u_{D}=u_{B}e^{\frac {1}{2}\kappa a}\cos \frac{1}{2}pa
\end{equation}
\begin{equation}\label{39}
  u_{B}=\left[ \frac {m}{m+E} \frac {1}{\kappa}\left(1+\cos pa\right) + \frac
{E+V_{0}}{E+V_{0}+m}a + \frac {m}{E+V_{0}+m}\frac {1}{p}\sin
pa\right]^{-\frac {1}{2}}
\end{equation}
\begin{equation}
    u_{C}=0
\end{equation}
and the energy spectrum satisfies
\begin{equation}
    \cot \frac {1}{2} pa= -\left[\frac {\left( m+E \right)\left( E+V_{0}-m \right)}{\left( m-E \right)\left( E+V_{0}+m
    \right)}\right]^{\frac {1}{2}}
\end{equation}

The upper component $u_1$ and the lower component $v_1$ of this
solution $\psi_{1, I}$ are even symmetric and odd symmetric
respectively. In another case, the coefficients of the alternative
solution $\psi_{1, II}$ in which the symmetry exchanges are given
by
\begin{equation}
  u_{A}=-u_{D}=-u_{C}e^{\frac {1}{2}\kappa a}\sin \frac{1}{2}pa
\end{equation}
\begin{equation}
  u_{B}=0
\end{equation}
\begin{equation}
    u_{C}=\left[ \frac {m}{m+E} \frac {1}{\kappa}\left( 1-\cos pa\right)+ \frac
{E+V_{0}}{E+V_{0}+m}a - \frac {m}{E+V_{0}+m}\frac {1}{p}\sin
pa\right]^{-\frac {1}{2}}
\end{equation}
and the energy spectrum satisfies
\begin{equation}
    \tan \frac {1}{2} pa= \left[\frac {\left( m+E \right)\left( E+V_{0}-m \right)}{\left( m-E \right)\left( E+V_{0}+m
    \right)}\right]^{\frac {1}{2}}.
\end{equation}

Another linearly independent solution $\psi_2$ can be obtained by
using Eq.~(\ref{96}). The result reads
\begin{equation}\label{5}
\begin{aligned}\psi_{2}=
\left( \begin{matrix} u_{2}\\[5pt]
v_{2}
\end{matrix}\right)=&\frac {1}{2}\left( \begin{matrix} -\frac {m+E}{\kappa}\\[5pt]
1
\end{matrix}\right)\frac {1}{u_{A}}e^{-\kappa y}\theta \left(-y-\frac {a}{2}\right)\\[5pt]
&+\Bigg [ \left( \begin{matrix} \frac {m+E+V_{0}}{p} \sin py\\[5pt]
\cos py
\end{matrix}\right)\frac {1}{u_{B}}+\left( \begin{matrix} -\frac {m+E+V_{0}}{p} \cos py\\[5pt]
\sin py
\end{matrix}\right)\frac {1}{u_{C}}\Bigg]\theta \left(-y+\frac {a}{2}\right)\theta \left(y+\frac
{a}{2}\right)\\[5pt]
&+\frac {1}{2}\left( \begin{matrix} \frac {m+E}{\kappa}\\[5pt]
1
\end{matrix}\right)\frac {1}{u_{D}}e^{\kappa y}\theta \left(y-\frac {a}{2}\right)
\end{aligned}\end{equation}
where $u_{A}, u_{B}, u_{C}$ and $u_{D}$ are given by Eqs.~(38-40)
for $\psi_{1, I}$ and Eqs.~(42-44) for $\psi_{1, II}$
respectively.
\subsection{Perturbation of the Square Well Potential}
Now we go to discuss the inhomogeneous equation (\ref{26}). The
one-well potential is given by
\begin{equation}
  \mathcal{V}=-V_{0}\left[ 1-\left( \frac {2y}{a}\right)^{2}\right]\theta \left( y+\frac {a}{2}
\right)\theta \left(-y+\frac {a}{2} \right)
\end{equation}
(Fig.~2) and the perturbation potential reads
\begin{equation}
  w\left( y \right)=\frac {4V_{0}}{a^{2}}y^{2}\theta \left( y+\frac {a}{2}
\right)\theta \left(-y+\frac {a}{2} \right)
\end{equation}
We calculate the first order energy shift
$\varepsilon^{\mathrm{(1)}}$. Using $\psi_1$ as the zeroth order
non-perturbation wave function $\psi^{(0)}$, then $ g^{(0)}
=\binom {1}{0}$ from Eq.~(\ref{34}). Substituting Eq.~(\ref{4})
and $g^{(0)}$ into Eq.~(\ref{1}), we obtain
\begin{equation}\label{6}
  \varepsilon^{(1)}_{I}=\frac
  {\displaystyle \frac {4V_{0}}{a^{2}}\left[ \frac {E+V_{0}}{E+V_{0}+m}\frac{1}{12}a^{3}+\frac {m}{E+V_{0}+m}\left(\frac {a^{2}}{4p}\sin pa+\frac {a}{2p^{2}}\cos pa-\frac {1}{2p^{3}}\sin pa\right)\right]}
  {\displaystyle\frac{m}{E+m}\frac{1}{\kappa }\left( 1+\cos
pa\right) +\left(
\frac{E+V_{0}}{E+V_{0}+m}a+\frac{m}{E+V_{0}+m}\frac{1}{p}\sin
pa\right) }
\end{equation}
for $\psi^{(0)}=\psi_{1,\ I}$, and
\begin{equation}\label{7}
  \varepsilon^{(1)}_{II}=\frac
  {\displaystyle \frac {4V_{0}}{a^{2}}\left[ \frac {E+V_{0}}{E+V_{0}+m}\frac{1}{12}a^{3}- \frac {m}{E+V_{0}+m}\left(\frac {a^{2}}{4p}\sin pa+\frac {a}{2p^{2}}\cos pa-\frac {1}{2p^{3}}\sin pa\right)\right]}
  {\displaystyle\frac{m}{E+m}\frac{1}{\kappa }\left( 1-\cos
pa\right) +\left(
\frac{E+V_{0}}{E+V_{0}+m}a-\frac{m}{E+V_{0}+m}\frac{1}{p}\sin
pa\right) }
\end{equation}
for $\psi^{(0)}=\psi_{1,\ II}$.

The wave function for one-well potential has been calculated up to
the first order iterative approximation. The results are
\begin{equation}\label{9}\begin{aligned}\psi^{(1)}_{I}=&
\left( \begin{matrix} u\\[5pt]
v
\end{matrix}\right)^{(1)}_{I}=\left( \begin{matrix} \displaystyle\frac{1}{\kappa} \left( -E y+\frac{m}{2\kappa}\right)\\[8pt]
\displaystyle-\frac {1}{m+E}\left(E y+\frac{m}{2\kappa}\right)
\end{matrix}\right)\varepsilon^{(1)}u_{A}e^{\kappa y}\theta \left(-y-\frac{a}{2}\right)+\\[9pt]
&+\left( \begin{matrix} \displaystyle\frac {1}{p}\left[\varepsilon^{(1)}\left(A\cos py+C\sin py\right)-\frac{4V_{0}}{a^{2}}\left(B\cos py+D\sin py\right)\right]\\[9pt]
\displaystyle\frac{1}{E+V_{0}+m}\left[\varepsilon^{(1)}\left(-A\sin
py+C\cos py\right)-\frac{4V_{0}}{a^{2}}\left(-B\sin py+D\cos
py\right)\right]
\end{matrix}\right)u_{B}\cdot\\[5pt]
&\theta \left(y+\frac {a}{2}\right)\theta \left(-y+\frac {a}{2}\right)+\left( \begin{matrix} \displaystyle\frac{1}{\kappa} \left(E y+\frac{m}{2\kappa}\right)\\[8pt]
\displaystyle\frac {1}{m+E}\left(-E y+\frac{m}{2\kappa}\right)
\end{matrix}\right)\varepsilon^{(1)}u_{D}e^{-\kappa y}\theta
\left(y-\frac{a}{2}\right)+C^{(1)}\psi_{1}
\end{aligned}\end{equation}
for $\psi^{(0)}=\psi_{1,\ I}$, and
\begin{equation}\label{10}\begin{aligned}\psi^{(1)}_{II}=&
\left( \begin{matrix} u\\[5pt]
v
\end{matrix}\right)^{(1)}_{II}=\left( \begin{matrix} \displaystyle\frac{1}{\kappa} \left( -E y+\frac{m}{2\kappa}\right)\\[8pt]
\displaystyle-\frac {1}{m+E}\left(E y+\frac{m}{2\kappa}\right)
\end{matrix}\right)\varepsilon^{(1)}u_{A}e^{\kappa y}\theta \left(-y-\frac{a}{2}\right)+\\[9pt]
&+\left( \begin{matrix} \displaystyle\frac {1}{p}\left[\varepsilon^{(1)}\left(A\cos py\mathbf{-}C\sin py\right)-\frac{4V_{0}}{a^{2}}\left(B\cos py\mathbf{-}D\sin py\right)\right]\\[9pt]
\displaystyle\frac{1}{E+V_{0}+m}\left[\varepsilon^{(1)}\left(\mathbf{+}A\sin
py+C\cos py\right)-\frac{4V_{0}}{a^{2}}\left(\mathbf{+}B\sin
py+D\cos py\right)\right]
\end{matrix}\right)u_{C}\cdot\\[5pt]
&\theta \left(y+\frac {a}{2}\right)\theta \left(-y+\frac {a}{2}\right)+\left( \begin{matrix} \displaystyle\frac{1}{\kappa} \left(E y+\frac{m}{2\kappa}\right)\\[8pt]
\displaystyle\frac {1}{m+E}\left(-E y+\frac{m}{2\kappa}\right)
\end{matrix}\right)\varepsilon^{(1)}u_{D}e^{-\kappa y}\theta
\left(y-\frac{a}{2}\right)+C^{(1)}\psi_{1}
\end{aligned}\end{equation}
for $\psi^{(0)}=\psi_{1,\ II}$ where
\begin{equation}
  A\equiv-\lambda\frac{m}{2p}\cos2py
\end{equation}
\begin{equation}
  B\equiv-\lambda\left(\frac{m}{2p}y^{2}\cos2py-\frac{m}{2p^{2}}y\sin2py-\frac{m}{4p^{3}}\cos2py\right)
\end{equation}
\begin{equation}
  C\equiv-\left[\left(E+V_{0}\right)y+\lambda\frac{m}{2p}\sin2py\right]
\end{equation}
\begin{equation}
  D\equiv-\left[\frac{1}{3}\left(E+V_{0}\right)y^{3}+\lambda\left(\frac{m}{2p}y^{2}\sin2py+\frac{m}{2p^{2}}y\cos2py-\frac{m}{4p^{3}}\sin2py\right)\right]
\end{equation}
and
\begin{equation}\lambda=\left\{\begin{aligned}
&+1\quad(\textnormal{for}\ \psi^{(0)}=\psi_{1,\ I})\\
&-1\quad(\textnormal{for}\ \psi^{(0)}=\psi_{1,\
II})\end{aligned}\right.
\end{equation}
$C^{(1)}$ is determined by normalization condition.
\pagebreak
\section{Double-Well Potential}%
The 1D Schr\"odinger equation with double-well potential has been
investigated by FLZ method~[3], WDA and variational method~[7]
previously. We now apply the EWDA to the double-well potential of
the 1D Dirac equation and give the energy eigenvalues to the first
order approximation.
\subsection{Double-Well and Double-Square-Well Potentials}
Given a symmetrical double-well potential
\begin{equation}\label{11}
 \mathcal{V}=\nu \left[ \left( \frac{y}{\eta}\right)^{2}-1\right]\left[ \left(
 \frac{y}{\xi}\right)^{2}-1\right]
\end{equation}
(Fig.~3) with the minimum of $\mathcal{V}$ being $-\mu$, define
\begin{equation}
  \tau \equiv \sqrt{\frac{\mu}{\nu}+1}+\sqrt{\frac{\mu}{\nu}}
\end{equation}
then we have
\begin{equation}
  \xi=\tau\eta
\end{equation}
With $\tau$ and $\eta$, instead of $\eta$ and $\xi$, $\mathcal{V}$
can be expressed as
\begin{equation}\label{12}
 \mathcal{V}=\nu \left[ \frac{1}{\tau^{2}}\left( \frac{y}{\eta}\right)^{4}-\left(1+\frac{1}{\tau^{2}}\right)\left(\frac{y}{\eta}\right)^{2}+1
\right]
\end{equation}

An exactly solvable double-square-well potential, whose bound
states can serve as the zeroth order wave function for iteration
procedure, is chosen in the homogeneous equation as,
\begin{equation}\label{13}\begin{aligned}
  V=&V_{0} \theta \left( -y-a \right)%
  -\mu \theta \left( y+a \right) \theta \left( -y-b \right)%
  +\nu \theta \left( y+b \right) \theta \left( -y+b \right)\\%
  &-\mu \theta \left( y-b \right) \theta \left( -y+a \right)%
  +V_{0} \theta \left( y-a \right)%
\end{aligned}\end{equation}
in which
\begin{equation}\label{14}
  V_{0}=\mathcal{V}\left( -a\right)=\mathcal{V}\left( a\right)
\end{equation}
where $a$ and $b$$\left( 0<b<a\right)$ are parameters which can be
adjusted to minimize the energy shift $\varepsilon^{(1)}$.

The wave function of a bound state solution to the Dirac equation
with the double-square-well potential is
\begin{equation}\label{16}
\begin{aligned}\psi_{1}=
\left( \begin{matrix} u_{1}\\[5pt]
v_{1}
\end{matrix}\right)=&+\left( \begin{matrix} \displaystyle 1\\[5pt]
\displaystyle \frac {m-E+V_{0}}{\kappa}
\end{matrix}\right)u_{A}e^{\kappa y}\theta \left(-y-a\right)\\[5pt]%
&+\left(\begin{matrix} \displaystyle u_{B}\cos \lambda y +\widetilde{u}_{B} \sin \lambda y \\[5pt]
\displaystyle \frac {m-E-\mu}{\lambda}\left( u_{B}\sin \lambda y -
\widetilde{u}_{B} \cos \lambda y\right)
\end{matrix}\right)%
\theta \left( y+b\right)\theta \left( -y-b\right)\\[5pt]%
&+\displaystyle\Bigg [ \left( \begin{matrix} \displaystyle \cosh qy\\[5pt]
\displaystyle\frac {m-E+\nu}{q} \sinh qy
\end{matrix}\right)u_{C}+\left( \begin{matrix} \displaystyle\sinh qy\\[5pt]
\displaystyle\frac {m-E+\nu}{q} \cosh qy
\end{matrix}\right)\widetilde{u}_{C}\displaystyle\Bigg]%
\theta \left(y+b\right)\theta \left(-y+b\right)\\[5pt]
&+\left(\begin{matrix} \displaystyle u_{D}\cos \lambda y +\widetilde{u}_{D} \sin \lambda y \\[5pt]
\displaystyle\frac {m-E-\mu}{\lambda}\left( u_{D}\sin \lambda y - \widetilde{u}_{D} \cos \lambda y\right)%
\end{matrix}\right)%
\theta \left( y-b\right)\theta \left( -y+a\right)\\[5pt]%
&+\left( \begin{matrix} \displaystyle1\\[5pt]
\displaystyle-\frac {m-E+V_{0}}{\kappa}
\end{matrix}\right)u_{E}e^{-\kappa y}\theta \left( y-a \right)
\end{aligned}\end{equation}
where
\begin{equation}
  \kappa = \sqrt{m^{2}-(E-V_{0})^{2}}\qquad(-m-V_{0}<E<m+V_{0})
\end{equation}
\begin{equation}
  \lambda = \sqrt{-m^{2}+ (E+\mu)^{2}}\qquad(E<-m-\mu\ or\ m-\mu<E)%
\end{equation}
\begin{equation}\label{}\label{17}
  q = \sqrt{m^{2}- (E-\nu)^{2}}\qquad(-m-\nu<E<m+\nu)
\end{equation}
If Eq.~(\ref{17}) dose not satisfy when $(E-\nu)^2>m^2$, we define
\begin{equation}\label{}\label{18}
  p = \sqrt{-m^{2}+ (E-\nu)^{2}}\qquad(E<-m-\mu\ or\ m+\mu<E)
\end{equation}
By replacing $q$ by $ip$ and $\widetilde{u}_{C}$ by
$-i\widetilde{u}_{C}$ the wave function (\ref{16}) and all the
following expressions maintain their forms.

Similar to the square-well case, the coefficients $u_{A}, u_{B},
\widetilde u_{B}, u_{C}, \widetilde u_{C}, u_{D}, \widetilde
u_{D}$ and $u_{E}$ of the solution $\psi_{1,\ I}$ with even
symmetric upper component $u_1$ and odd symmetric lower component
$v_1$ are given by
\begin{equation}
  u_{B}=u_{D}=\alpha u_{C}
\end{equation}
\begin{equation}
  \widetilde{u}_{B}=-\widetilde{u}_{D}=-\widetilde{\alpha} u_{C}
\end{equation}
\begin{equation}
  u_{A}=u_{E}=\gamma u_{C}
\end{equation}
\begin{equation}\begin{aligned}
  u_{C}=&\left[\frac{4m}{m+E-V_{0}}\gamma^{2}C+\frac{4m}{m+E+\mu}%
  \left( \alpha^{2}-\widetilde{\alpha}^{2}\right)F+\frac{4\left(E+\mu\right)}{m+E+\mu}%
  \left( \alpha^{2}+\widetilde{\alpha}^{2}\right)\frac{1}{2}\left(%
  a-b\right)\right.\\
  &\left.-\frac{4m}{m+E+\mu}\alpha\widetilde{\alpha}L+\frac{2m}{m+E-\nu}%
  \frac{1}{2q}\sinh 2qb +
  \frac{2\left(E-\nu\right)}{m+E-\nu}b\right]^{-\frac{1}{2}}
\end{aligned}\end{equation}
and the energy spectrum satisfies
\begin{equation}\label{19}
  \tanh qb = \frac%
  {\displaystyle\frac{\kappa}{m-E+V_{0}}\frac{m-E-\mu}{\lambda}\tan\lambda\left(b-a\right)-1}%
  {\displaystyle\left(\frac{\lambda}{m-E-\mu}\tan
  \lambda\left(b-a\right)+\frac{\kappa}{m-E+V_{0}}\right)\frac{m-E+\mu}{q}}.
\end{equation}

Correspondingly, the coefficients of the alternative solution
$\psi_{1,\ II}$ with the exchanged symmetry of components are
given by
\begin{equation}
  u_{B}=-u_{D}=-\beta \widetilde{u}_{C}
\end{equation}
\begin{equation}
  \widetilde{u}_{B}=\widetilde{u}_{D}=\widetilde{\beta}\widetilde{u}_{C}
\end{equation}
\begin{equation}
  u_{A}=-u_{E}=-\delta \widetilde{u}_{C}
\end{equation}
\begin{equation}\begin{aligned}
  \widetilde{u}_{C}=&\left[\frac{4m}{m+E-V_{0}}\delta^{2}C+\frac{4m}{m+E+\mu}%
  \left( \beta^{2}-\widetilde{\beta}^{2}\right)F+\frac{4\left(E+\mu\right)}{m+E+\mu}%
  \left( \beta^{2}+\widetilde{\beta}^{2}\right)\frac{1}{2}\left(%
  a-b\right)\right.\\
  &\left.-\frac{4m}{m+E+\mu}\beta\widetilde{\beta}L+\frac{2m}{m+E-\nu}%
  \frac{1}{2q}\sinh 2qb -
  \frac{2\left(E-\nu\right)}{m+E-\nu}b\right]^{-\frac{1}{2}}
\end{aligned}\end{equation}
and the energy spectrum satisfies
\begin{equation}
  \coth qb = \frac%
  {\displaystyle\frac{\kappa}{m-E+V_{0}}\frac{m-E-\mu}{\lambda}\tan\lambda\left(b-a\right)-1}%
  {\displaystyle\left(\frac{\lambda}{m-E-\mu}\tan
  \lambda\left(b-a\right)+\frac{\kappa}{m-E+V_{0}}\right)\frac{m-E+\mu}{q}}
\end{equation}

In the above expressions we have defined
\begin{equation}
  \alpha = \cosh qb \cos \lambda b%
  + \frac{\lambda}{m-E-\mu}\frac{m-E+\nu}{q}\sinh qb \sin \lambda b%
\end{equation}
\begin{equation}
  \widetilde{\alpha} = \cosh qb \sin \lambda b%
  - \frac{\lambda}{m-E-\mu}\frac{m-E+\nu}{q}\sinh qb \cos \lambda b%
\end{equation}
\begin{equation}
  \beta = \sinh qb \cos \lambda b%
  + \frac{\lambda}{m-E-\mu}\frac{m-E+\nu}{q}\cosh qb \sin \lambda b%
\end{equation}
\begin{equation}
  \widetilde{\beta} = \sinh qb \sin \lambda b%
  - \frac{\lambda}{m-E-\mu}\frac{m-E+\nu}{q}\cosh qb \cos \lambda b%
\end{equation}
\begin{equation}
  \gamma =e^{\kappa a}\left[ \cosh qb \cos \lambda(b-a)%
  + \frac{\lambda}{m-E-\mu}\frac{m-E+\nu}{q}\sinh qb \cos \lambda (b-a)\right]%
\end{equation}
\begin{equation}
  \delta =e^{\kappa a}\left[ \sinh qb \cos \lambda(b-a)%
  + \frac{\lambda}{m-E-\mu}\frac{m-E+\nu}{q}\cosh qb \sin \lambda (b-a)\right]%
\end{equation}
and $C$, $F$ and $L$ given later in Eqs.~(90), (93) and (96).

Another linearly independent solution, though it is not necessary
for deriving the first order energy shift. But it is involved in
higher order calculations. The result, which can be constructed by
Eq.~(\ref{96}), reads
\begin{equation}
\begin{aligned}\psi_{2}=
\left( \begin{matrix} u_{2}\\[5pt]
v_{2}
\end{matrix}\right)=&+\frac{1}{2}\left( \begin{matrix} \displaystyle \frac {-m-E+V_{0}}{\kappa}\\[5pt]
\displaystyle 1
\end{matrix}\right)\frac{1}{u_{A}}e^{-\kappa y}\theta \left(-y-a\right)\\[5pt]%
&+\left(\begin{matrix} \displaystyle \frac{m+E+\mu}{\lambda}\left(u_{B}\sin\lambda y -\widetilde{u}_{B} \cos \lambda y \right)\\[5pt]
\displaystyle \left( u_{B}\cos \lambda y + \widetilde{u}_{B} \sin
\lambda y\right)
\end{matrix}\right)\frac{1}{u_{B}^{2}+\widetilde{u}_{B}^{2}}%
\theta \left( y+a\right)\theta \left( -y-b\right)\\[5pt]%
&+\displaystyle\Bigg [ \left( \begin{matrix} \displaystyle \frac {m+E-\nu}{q} \sinh qy\\[5pt]
\displaystyle\cosh qy
\end{matrix}\right)\frac{1}{u_{C}}+\left( \begin{matrix} \displaystyle-\frac {m+E-\nu}{q} \cosh qy\\[5pt]
\displaystyle-\sinh qy
\end{matrix}\right)\frac{1}{\widetilde{u}_{C}}\displaystyle\Bigg]%
\theta \left(y+b\right)\theta \left(-y+b\right)\\[5pt]
&+\left(\begin{matrix} \displaystyle \frac {m+E+\mu}{\lambda}\left( u_{D}\sin \lambda y - \widetilde{u}_{D} \cos \lambda y\right)\\[5pt]
\displaystyle u_{D}\cos \lambda y +\widetilde{u}_{D} \sin \lambda y %
\end{matrix}\right)\frac{1}{u_{D}^{2}+\widetilde{u}_{D}^{2}}%
\theta \left( y-b\right)\theta \left( -y+a\right)\\[5pt]%
&+\frac{1}{2}\left( \begin{matrix} \displaystyle \frac {m+E-V_{0}}{\kappa}\\[5pt]
\displaystyle 1
\end{matrix}\right)\frac{1}{u_{E}}e^{\kappa y}\theta \left( y-a \right)
\end{aligned}\end{equation}
where $u_{A}, u_{B}, \widetilde u_{B}, u_{C}, \widetilde u_{C},
u_{D}, \widetilde u_{D}$ and $u_{E}$ are given by Eqs.~(69-72) for
$\psi_{2,\ I}$ and Eqs.~(74-77) for $\psi_{2,\ II}$ respectively.
\subsection{First Order Energy Shift for the Double-Well Potential}
The perturbation potential on the double-square-well potential
reads
\begin{equation}
  w(y)=\mathcal{V}(y)-V(y)
\end{equation}

Using $\psi_1$ as the zeroth order $\psi^{(0)}$ in iterative
procedure, and $g^{(0)}=\binom{1}{0}$ from Eq.~(\ref{34}), the
first order energy shift $\varepsilon^{(1)}$ is obtained by
substituting Eq.~(\ref{16}) and $g^{(0)}$ into Eq.~(\ref{1}), we
get
\begin{equation}\label{50}\begin{aligned}
  &\varepsilon^{(1)}=\\[5pt]
  &+\frac{4m}{m+E-V_{0}}u_{A}^{2}\left[%
  \frac{\nu}{\tau^{2}\eta^{4}}A-\left(1+\frac{1}{\tau^{2}}\right)%
  \frac{\nu}{\eta^{2}}B+\left(\nu-V_{0}\right)C\right]\\[5pt]
  &+\frac{4m}{m+E+\mu}\left(u_{B}^{2}-\widetilde{u}_{B}^{2}\right)\left[%
  \frac{\nu}{\tau^{2}\eta^{4}}D-\left(1+\frac{1}{\tau^{2}}\right)%
  \frac{\nu}{\eta^{2}}G+\left(\nu+\mu\right)F\right]\\[5pt]
  &+\frac{4(E+\mu)}{m+E+\mu}\left(u_{B}^{2}+\widetilde{u}_{B}^{2}\right)\left[%
  \frac{\nu}{\tau^{2}\eta^{4}}\frac{1}{10}(a^{5}-b^{5})-\left(1+\frac{1}{\tau^{2}}\right)%
  \frac{\nu}{\eta^{2}}\frac{1}{6}(a^{3}-b^{3})+\left(\nu+\mu\right)\frac{1}{2}(a-b)\right]\\[5pt]
  &+\frac{4m}{m+E+\mu}u_{B}\widetilde{u}_{B}\left[%
  \frac{\nu}{\tau^{2}\eta^{4}}J-\left(1+\frac{1}{\tau^{2}}\right)%
  \frac{\nu}{\eta^{2}}K+\left(\nu+\mu\right)L\right]\\[5pt]
  &+\frac{2m}{m+E-\nu}\left(u_{C}^{2}+\widetilde{u}_{C}^{2}\right)\left[%
  \frac{\nu}{\tau^{2}\eta^{4}}M-\left(1+\frac{1}{\tau^{2}}\right)%
  \frac{\nu}{\eta^{2}}N\right]\\[5pt]
  &+\frac{2(E-\nu)}{m+E-\nu}\left(u_{C}^{2}-\widetilde{u}_{C}^{2}\right)\left[%
  \frac{\nu}{\tau^{2}\eta^{4}}\frac{1}{5}b^{5}-\left(1+\frac{1}{\tau^{2}}\right)%
  \frac{\nu}{\eta^{2}}\frac{1}{3}b^{3}\right],
\end{aligned}\end{equation}
where
\begin{equation}
  A=\left( \frac{a^{4}}{2\kappa} + \frac{a^{3}}{\kappa^{2}}%
  + \frac{3a^{2}}{2\kappa^{3}} + \frac{3a}{2\kappa^{4}}%
  +\frac{3}{4\kappa^{5}} \right)e^{-2\kappa a}
\end{equation}
\begin{equation}
  B=\frac{1}{2}\left( \frac{a^{2}}{\kappa} + \frac{a}{\kappa^{2}}%
  + \frac{1}{2\kappa^{3}}\right)e^{-2\kappa a}
\end{equation}
\begin{equation}
  C=\frac{1}{2\kappa}e^{-2\kappa a}
\end{equation}
\begin{equation}\begin{aligned}
  D=&+\frac{1}{4}\left( \frac{a^{4}}{\lambda}-\frac{3a^{2}}{\lambda^{3}}%
  +\frac{3}{2\lambda^{5}}\right)\sin 2\lambda a+ \frac{1}{4}\left(%
  \frac{2a^{3}}{\lambda^{2}}-\frac{3a}{2\lambda^{4}}\right)%
  \cos 2\lambda a\\
  &-\frac{1}{4}\left( \frac{b^{4}}{\lambda}-\frac{3b^{2}}{\lambda^{3}}%
  -\frac{3}{2\lambda^{5}}\right)\sin 2\lambda b~- \frac{1}{4}\left(%
  \frac{2b^{3}}{\lambda^{2}}-\frac{3b}{2\lambda^{4}}\right)%
  \cos 2\lambda b
\end{aligned}\end{equation}
\begin{equation}\begin{aligned}
 G=&+\frac{1}{4} \left(\frac{a^{2}}{\lambda} -
 \frac{1}{2\lambda^{3}}\right)\sin 2\lambda a %
 +\frac{a}{4 \lambda^{2}}\cos 2\lambda a\\%
 &-\frac{1}{4} \left(\frac{b^{2}}{\lambda} -
 \frac{1}{4\lambda^{3}}\right)\sin 2\lambda b %
 -\frac{b}{ \lambda^{2}}\cos 2\lambda b
\end{aligned}\end{equation}
\begin{equation}
  F=\frac{1}{4\lambda}\left(\sin 2\lambda a -\sin 2\lambda b\right)
\end{equation}
\begin{equation}\begin{aligned}
  J=&+\frac{1}{2}\left( \frac{a^{4}}{\lambda}-\frac{3a^{2}}{\lambda^{3}}%
  +\frac{3}{2\lambda^{5}}\right)\cos 2\lambda a- \frac{1}{2}\left(%
  \frac{2a^{3}}{\lambda^{2}}-\frac{3a}{\lambda^{4}}\right)%
  \sin 2\lambda a\\
  &-\frac{1}{2}\left( \frac{b^{4}}{\lambda}-\frac{3b^{2}}{\lambda^{3}}%
  +\frac{3}{2\lambda^{5}}\right)\cos 2\lambda b~+ \frac{1}{2}\left(%
  \frac{2b^{3}}{\lambda^{2}}-\frac{3b}{\lambda^{4}}\right)%
  \sin 2\lambda b
\end{aligned}\end{equation}
\begin{equation}\begin{aligned}
 K=&+\frac{1}{2} \left(\frac{a^{2}}{\lambda} -
 \frac{1}{2\lambda^{3}}\right)\cos 2\lambda a %
 -\frac{a}{2 \lambda^{2}}\sin 2\lambda a\\%
 &-\frac{1}{2} \left(\frac{b^{2}}{\lambda} -
 \frac{1}{2\lambda^{3}}\right)\cos 2\lambda b %
 +\frac{b}{2 \lambda^{2}}\sin 2\lambda b
\end{aligned}\end{equation}
\begin{equation}
  L=\frac{1}{2\lambda}\left(\cos 2\lambda a -\cos 2\lambda b\right)
\end{equation}
\begin{equation}
  M=\frac{1}{2}\left(\frac{b^{4}}{q}+\frac{3b^{3}}{q^{3}}+\frac{3}{2q^{5}}\right)%
  \sinh 2qb - \frac{1}{2}\left( \frac{2b^{3}}{q^{2}}+\frac{3b}{2q^{4}}\right)%
  \cosh 2qb
\end{equation}
\begin{equation}
  N=\frac{1}{2}\left(\frac{b^{2}}{q}+\frac{1}{2q^{3}}\right)%
  \sinh 2qb - \frac{b}{2q^{2}}\cosh 2qb
\end{equation}
\pagebreak
\section{Discussion and Conclusion}

In summary, a new method, namely, the Extended Wronskian
Determinant Approach has been suggested to calculate the energy
spectra and the wave functions of 1D Dirac equation. An integral
equation which can be solved by iterative procedure has been
established to study the inhomogeneous Dirac equation with
perturbation potential. Using the solution of the corresponding
homogeneous Dirac equation as the zeroth wave function, we can
calculate the wave function and energy spectrum of the
perturbative Dirac equation up to any order in principle. Tow
kinds of exact solutions including solution of one square-well
potential and double-square-well potential have been employed as
the zeroth approximations to study the one-well potential and
double-well potential respectively. For one-well case, the first
order iterative approximation to the energy levels and the wave
functions has been obtained. For double-well case, we have got the
energy level approximation up to the first order. These two
examples confirm that the EWDA is a successful method to
investigate the 1D Dirac equation.

In fact, it is not necessary to adopt the exact solutions of
homogeneous 1D Dirac equation as the zeroth order wave functions.
Instead of exact solution, we ca use a trial wave function. And we
can use the variational method for the non-perturbative
Hamiltonian to determine the best choice of the parameters in the
trial wave function first, and then use the trial wave function
with the best parameters as the zeroth order wave function to
study the perturbative Dirac equation by EWDA. \linebreak[16]
\pagebreak
\begin{center}
\textbf{\large Acknowledgments}
\end{center}

We thank J.~Qiu and Prof. W.~Q.~Zhao for helpful discussions. This
work was supported in part by National Natural Science Foundation
of China under Grant NOS. 10047005, 10235030, 19947001.
\pagebreak

\pagebreak
\begin{center}
\textbf{\large Figure Captions} %
\end{center}

Figure 1: one-well potential%

Figure 2: perturbed one-well potential%

Figure 3: square-double-well and double-well potentials%
\end{document}